# Microscopic Theory of the Influence of Strong Attractive Forces on the Activated Dynamics of Dense Glass and Gel Forming Fluids

Ashesh Ghosh[1,3] and Kenneth S. Schweizer [1-4,*]

[1]Department of Chemistry, [2]Department of Material Science, [3]Materials Research Laboratory, [4]Department of Chemical & Biomolecular Engineering
University of Illinois @ Urbana-Champaign, Illinois 61801, USA
*email: kschweiz@illinois.edu

**Abstract**
We theoretically study the non-monotonic (re-entrant) activated dynamics associated with a repulsive glass to fluid to attractive glass transition in high density particle suspensions interacting via strong short-range attractive forces. The classic theoretical "projection" approximation that replaces all microscopic forces by a single effective force determined solely by equilibrium pair correlations is revisited based on the "projectionless dynamic theory" (PDT) that avoids force projection. A hybrid-PDT is formulated that explicitly quantifies how attractive forces induce dynamical constraints, while singular hard core interactions are treated based on the projection approach. Both the effects of interference between repulsive and attractive forces, and structural changes due to attraction-induced bond formation that competes with caging, are included. Combined with the microscopic Elastically Collective Nonlinear Langevin Equation (ECNLE) theory of activated relaxation, the resultant approach appears to properly capture both the re-entrant dynamic crossover behavior and the strong non-monotonic variation of the activated structural relaxation time with attraction strength and range at very high volume fractions. Qualitative differences with ECNLE theory-based results that adopt the full projection approximation are identified, and testable predictions made. The new formulation appears qualitatively consistent with multiple experimental and simulation studies, and provides a new perspective for the overall problem that is rooted in activated motion and interference between repulsive and attractive forces. This is conceptually distinct from empirical shifting or other ad hoc modifications of ideal mode coupling theory which do not take into account activated dynamics. Implications for thermal glass forming liquids are briefly discussed.

## I. Introduction

Understanding glassy dynamics and kinetic arrest remains a multi-faceted grand challenge in statistical mechanics, condensed matter physics, and materials science [1-4]. The simplest system is the hard sphere (HS) fluid and its experimental realization as a colloidal suspension characterized by a single dimensionless parameter, the volume fraction $\phi = \frac{\pi}{6} \rho \sigma^3$, where $\rho$ is the number density and $\sigma$ the particle diameter. Beyond the equilibrium crystallization volume fraction (~0.495), the structural relaxation time grows very quickly with density [2,4,5]. In simulation and experiment a practical kinetic vitrification transition occurs when the structural or alpha relaxation time scale exceeds the maximum value measurable. In practice, this typically occurs at $\phi_g \sim 0.58 - 0.6$ [5,6]. But even well below this volume fraction, examination of trajectories shows the strong growth of the relaxation time and viscosity is accompanied by long periods of transient particle localization in a cage, punctuated with large amplitude activated hopping events characteristic of intermittent dynamics [5,7-11].

The earliest microscopic force level theory for hard sphere dynamics is the ideal mode coupling theory (MCT) [3]. It is based on the idea that local caging and nonlinearly coupled collective density fluctuations induce large slowing down at high enough density resulting in strict solidification (infinite relaxation time). The relaxation time, $\tau_\alpha(\phi)$, is predicted to diverge as an inverse critical power law as $\phi \to \phi_c \sim 0.515$, which is far below the practical experimental or simulation vitrification volume fraction. Recent Generalized Mode Coupling Theory (GMCT) efforts [12-15] include some aspects of higher order correlation functions ignored in classic ideal MCT, but still invoke a self-consistent dynamic closure. The latest work finds the critical volume fraction at the most complex level of the theory shifts to $\phi_c = 0.56$ [15]. However, GMCT still predicts a strict critical power law divergence of the structural relaxation time with an exponent only modestly larger than the classic ideal MCT value of ~2.5 [15]. Hence, activated processes as commonly envisioned to underlie an exponential variation of a time scale with the relevant control variable are not captured, which presumably requires an "infinite order" treatment within the ensemble-averaged GMCT framework [16].

A critical power law form of the relaxation time per the original ideal MCT does agree with experiments and simulations on hard sphere fluids and suspensions over ~3 decades, but only *if* $\phi_c$ is empirically *shifted* to higher values [3, 5,17,18]. At still higher volume fractions, both colloid experiments and computer simulations find the critical power law qualitatively fails, and the time scale grows in an exponential or activated manner [19]. This phenomenology suggests ideal MCT might describe the initial stages of slowing down, but the divergence is an artifact of not capturing activated processes. However, since nonperturbative signatures of strongly non-Gaussian intermittent motion [5,8,9,11,20] (e.g., large non-Gaussian factors, van Hove function with exponential tails) and direct visualization of hopping trajectories emerge well below $\phi \sim 0.58$, the physical significance of empirical shifting of the MCT critical volume fraction to compare with experiment or simulation is conceptually unclear.

The above considerations motivated the development of a force level microscopic approach which adopts some MCT ideas for quantifying dynamical caging constraints via equilibrium pair correlations, but goes beyond it to include thermal noise driven activated hopping at the single particle level -- the Nonlinear Langevin Equation (NLE) theory [10,11,20,21]. Most recently this approach was extended to include coupling of cage scale hopping with coordinated collective elastic fluctuations outside the cage, the Elastically Collective NLE (ECNLE) theory [22-24]. ECNLE theory captures well the observable ~5 decades of hard sphere fluid glassy relaxation.

A qualitatively different type of dynamically arrested state is obtained for particle fluids with strong short range attractions [6,25-33]. They can be experimentally realized by adding small non-adsorbing polymers to colloidal suspensions, or by introducing thermosensitive brushes on particle surfaces. In the former case, depletion attraction is controlled by polymer concentration (strength) and radius of gyration $R_g$ (spatial range). Of interest here is when $R_g \ll R$ (where $R$ is the colloid radius) corresponding to a short range attraction which can strongly modify colloid packing and induce physical clustering. This type of attraction cannot be realized in typical nonpolar thermal liquids where attractions are much longer range (e.g., Lennard-Jones potential (LJ)) and repulsive interactions dominate structure [34-36].

Introducing short range attraction raises in the possibility of physical "bond" formation, which competes with caging due to repulsive interactions. In experiment [25-29], on the computer [27,29,32], and in ideal MCT [3,25,27,32,33], if the attraction is sufficiently weak and short range, relaxation can strongly speed up and practical vitrification is delayed to higher volume fraction or lower temperature ("glass melting" or "re-entrancy"). However, with further increasing of the attraction strength an "attractive glass" (AG) state emerges characterized by both bonding and caging which reinforce, resulting in a relaxation time that can far exceed that of the hard sphere fluid. This phenomenology is qualitatively captured at the ideal MCT level via the amplitude of the collective static structure factor at its cage peak, $S(k^*)$, evolving in a non-monotonic manner with attraction strength [3,32,33]. This cage coherence effect controls slow dynamics despite the contact value $(g(r = \sigma))$ of the pair correlation function monotonically increasing with attraction strength. However, invoking MCT to explain the experimental and simulation behavior seems problematic since relaxation times generally do not grow as inverse critical power laws, trajectories are highly intermittent implicating non-Gaussian processes, and a rather large empirical shift of volume fraction is required to align the ideal nonergodicity boundary with the much higher volume fractions where the new phenomena are observed in experiment and simulation. Figure 1 sketches the situation of interest in this article.

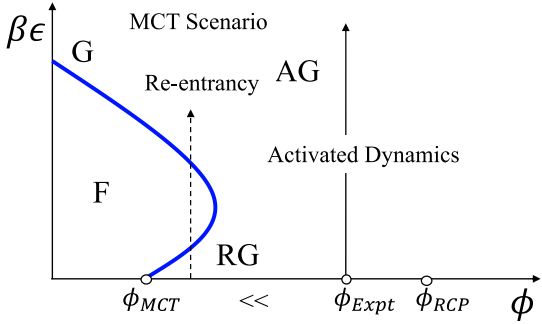

Figure 1: Schematic of the re-entrant glass melting kinetic arrest diagram suggested by ideal MCT, simulations, and experiments. 'F' is an ergodic fluid state, 'RG' is a repulsive glass dominated by caging constraints, 'AG' is an Attractive Glass controlled by both caging and bonding forces. At lower volume fractions, a dense gel state, 'G', can exist. Experiments and simulations find this phenomenology emerges only for volume fractions well beyond (~0.56-0.61) that predicted by classic ideal MCT, in a regime where intermittent activated dynamics at the trajectory and ensemble-averaged levels is important.

To set the stage for our work, we first recall the key approximations of classic ideal MCT [3,37] which focuses on ensemble averaged time correlation functions such as the collective density fluctuation $S(k,t)$ and its single particle analog $F_s(k,t)$. First, slow dynamics is assumed to be slaved to structural correlations, an idea implemented by "projecting" the real Newtonian pair decomposable forces onto equilibrium pair correlations. Any consequences of *forces* on dynamics beyond the influence of the corresponding *potential* on pair structure is not retained. Second, a required four-point (in space-time) dynamic density fluctuation correlation function (Figure 2) is factorized in a Gaussian fashion to close the theory at the level of two-point correlators. This critical approximation precludes capturing activated processes which are extremely non-Gaussian, and leads to the unphysical critical power law divergence of the relaxation time. Addressing the second issue has been the focus of the work by Schweizer and coworkers to develop microscopic activated dynamics theories (NLE, ENCLE) based on retaining the projection approximation [21,23]. Our goal here is to go beyond the latter approximation. The question of two liquid systems with very similar pair structure but different interparticle forces having strongly different dynamics (e.g., WCA vs LJ) for some thermodynamic states has been studied via simulation in the weakly supercooled regime for simple spherical particle mixture models [38-41]. A theoretical effort to address those findings was carried out by Dell and Schweizer [42] where the strict projection was avoided (projectionless dynamic theory, PDT) and activated hopping included. However, our focus here is on the different soft matter systems which interact via strong short range attractions, and how an activated dynamics approach that explicitly takes into account the corresponding strong attractive *forces* can be constructed.

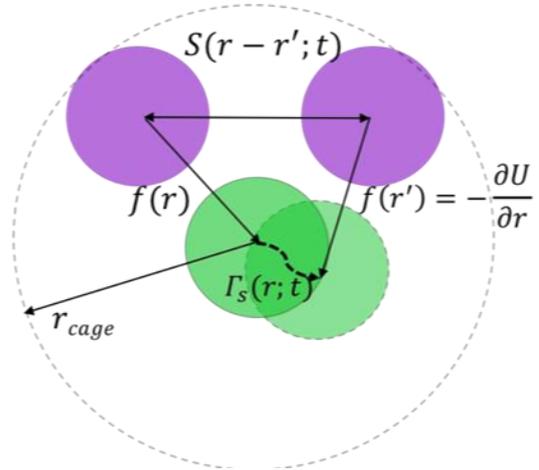

Figure 2: Schematic of the required 4 point (3 different particles) force-force time correlation in naïve MCT. Dynamic relaxation occurs via tagged particle and collective motion. Forces are either effective (projection approximation) or bare Newtonian (PDT).

Section II briefly recalls the single particle naïve MCT (NMCT), NLE theory and ECNLE theory based on the projection approximation. A hybrid projectionless dynamics analog of all the latter theories is developed in section III, and general aspects discussed. Numerical results based on the new approach for ideal localization boundaries, transient localization length, barriers, and activated relaxation times are presented in section IV. These results are contrasted with those obtained based on the projection-based version of ECNLE theory. Our new results are contrasted with experiments and simulations in Section V. The article concludes in section VI with a discussion. Supplementary Materials (SM) includes details of the force vertex analysis, and results for thermal liquids with relatively long range attractive forces.

## II. Theoretical Background

We recall the NMCT, NLE and ECNLE theories for hard spheres based on the force projection approximation. All details are discussed in many prior articles [10,21-24].

### A. Ideal NMCT of Spherical Particle Liquids

The crucial quantity is the force-force time correlation function,

$$K(t) = \frac{\beta}{3}\langle \vec{F}_0(0) \cdot \vec{F}_0(t) \rangle \quad (1)$$

where $\vec{F}_0$ is the total force on a tagged particle (indicated by 0) due to all surrounding particles. Force relaxation is assumed to be strongly coupled to slow modes identified (per MCT) as the product of the tagged particle density and collective density. Projecting the forces on this slow mode, followed by factorizing 4-point correlations into a product of 2-point ones, and replacing projected dynamics with the real full dynamics, one obtains in Fourier space [10,43],

$$K(t) \quad (2)$$
$$= \frac{\beta\rho}{3} \int \frac{d\vec{k}}{(2\pi)^3} k^2 C(k)^2 S(k) \Gamma_s(k,t) \Gamma_c(k,t)$$

where $\beta = (k_B T)^{-1}$, $S(k) = (1 - \rho C(k))^{-1}$ is the static structure factor, $C$ is the direct correlation function, and $\Gamma_s(k,t) = \langle e^{i\vec{k}\cdot(\vec{r}(t)-\vec{r}(0))} \rangle$ and $\Gamma_c(k,t) = S(k,t)/S(k)$ are the normalized (at $t=0$) single and collective dynamic propagators, respectively. The projection approach effectively replaces *all* real forces by the gradient of the direct correlation function:

$$\vec{F}(r) = k_B T \vec{\nabla} C(r) \quad (3)$$

The force time correlations on a tagged particle decay via two parallel channels, collective motion of the surrounding media and tagged particle motion. The putative kinetically arrested state is taken to be an Einstein glass with localization length scale $r_L$. In an arrested state the propagators become Debye-Waller factors [10, 43], and analyzing the long time limit one obtains

$$\Gamma_s(k, t \to \infty) = e^{-\frac{k^2 r_L^2}{6}} \quad (4)$$

$$\Gamma_c(k, t \to \infty) = e^{-\frac{k^2 r_L^2}{6 S(k)}} \quad (5)$$

$$\frac{1}{r_L^2} = \frac{1}{9} \int \frac{d\vec{k}}{(2\pi)^3} \rho k^2 C(k)^2 S(k) e^{-\frac{k^2 r^2}{6}[1+S^{-1}(k)]} \quad (6)$$

Equation (6) is the self-consistent ideal localization relation of NMCT [10].

Equation (6) predicts for hard spheres a literal dynamical arrest transition at $\phi_c \approx 0.432$ using Percus-Yevick (PY) theory for the input structure [10]. If there is a short range attraction, NMCT predicts an ideal nonergodicity boundary [44,45] qualitatively like in Figure 1 with a re-entrant nose feature, with a dependence on attraction strength and range qualitatively the same as for full ideal MCT [27,32,33]. Post-facto analysis of NMCT results [45-47] reveals the same origin for the re-entrant or glass melting behavior as in the full MCT: a non-monotonic variation with attraction strength of the effective mean square force vertex due to changes of $S(k \sim k^*)$ (see SM). The latter is defined per Eq(6) as $V(k) = \rho^{-1} k^4 (1 - S^{-1}(k))^2 S(k))$. The ideal MCT transition signals a crossover to activated motion.

### B. Activated Dynamics Theories

Nonlinear Langevin Equation (NLE) theory is a microscopic approach for single particle stochastic trajectories which captures thermal fluctuation driven activated hopping [10, 21]. It was heuristically constructed based on a combination of physical arguments, elements of MCT to quantify dynamical constraints, and a *coarse-grained* version of dynamic density functional theory [21]. NLE theory is *not* formulated in terms of ensemble-averaged time correlation functions and includes trajectory fluctuations, which is the reason it can capture the nonperturbative activated barrier hopping event.

The central new concept is a "dynamic free energy", $F_{dyn}(r)$, the gradient of which is the systematic force that controls single particle trajectories at the level of an angularly averaged scalar displacement variable $r(t)$. The NLE equation in the overdamped limit is [21],

$$-\zeta_s \frac{dr(t)}{dt} - \frac{\partial F_{dyn}(r(t))}{\partial r(t)} + \delta f(t) = 0 \quad (7)$$

where the thermal noise term satisfies, $\langle \delta f(0)\delta f(t)\rangle = 2k_B T \zeta_s \delta(t)$, $r(t=0)=0$, and the first term describes a "short time" non-activated frictional process. In this article, time will be expressed in terms of the corresponding short time scale, $\tau_s = \beta \zeta_s \sigma^2$ (an explicit expression for spheres given elsewhere [21]). The key quantity is the dynamic free energy,

$$\beta F_{dyn}(r) = -3\ln(r) - \frac{\rho}{2\pi^2}\int_0^\infty \frac{|M(k)|^2 S(k)}{1+S^{-1}(k)} e^{-\frac{k^2 r^2}{6}(1+S^{-1}(k))} dk \quad (8)$$

where a dynamic force vertex, $|M(k)|$, is defined. Within the projection approach one has:

$$\vec{M}_{NMCT}(k) = kC(k)\hat{k} \quad (9)$$

The first term in Eq(8) favors the fluid state and second term favors a localized state. If the gradient of the dynamic free energy is set to zero, or if the noise in the NLE is dropped, particles arrest at a displacement given by the NMCT localization Eq(6). For $\phi > \phi_c$, the dynamic free energy acquires a minimum at $r = r_L$ and a barrier of height $F_B$ at $r = r_B$ (see Figure 3).

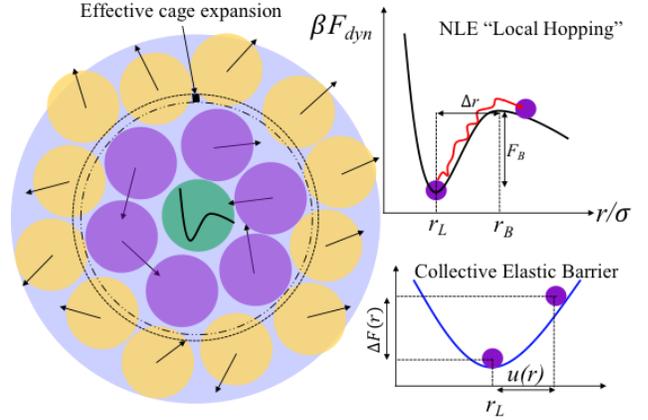

Figure 3: Left: schematic of the physical content of ECNLE theory involving cage scale hopping coupled to a longer range collective elastic fluctuation. The amplitude of the elastic displacements outside the cage are highly exaggerated in the shown image, and in practice are predicted to be of order the transient localization length or smaller. Right: schematic of the dynamic free energy with its key length and energy scales defined; beyond the cage scale a harmonic displacement field characterizes the collective elastic fluctuation.

Motivated by ideas in phenomenological elastic models of glassy dynamics [48,49], NLE theory has recently been generalized to take into account coupling of local cage scale hopping and collective elastic distortion of the particles outside of the cage, yielding ECNLE theory [22-24]. The elastic barrier is given by $\beta F_{el} = 4\pi \int_{r_{cage}}^\infty r^2 \rho g(r) \left(\frac{1}{2} K_0 u(r)^2\right) dr$, where $K_0$ is the harmonic spring constant of the dynamic free energy at its minimum, $u(r)$ is the elastic displacement field given by $u(r) = \frac{3\Delta r^2}{32 r_{cage}}\left(\frac{r_{cage}}{r}\right)^2$, $r > r_{cage}$ where $r$ is the distance from the cage center, and $\Delta r = r_B - r_L$ is the jump distance. The elastic barrier scales as the product of volume fraction, spring constant $K_0$ (sets the energy scale), and 4 powers of the jump distance. All information needed to compute the elastic barrier follows from the dynamic free energy.

In ECNLE theory the alpha relaxation is a mixed local-nonlocal activated event with a total barrier equal to the sum of the cage and longer range collective elastic contributions, $F_{total} = F_B + F_{el}$. The non-dimensionalized mean structural relaxation time is taken to be the mean barrier hopping time per Kramers theory [23],

$$\frac{\tau_\alpha}{\tau_s} =$$
$$= \sigma^{-2} e^{\beta F_{el}}$$
$$\times \int_{r_L}^{r_B} dx\, e^{\beta F_{dyn}(x)} \int_{r_L}^{x} dy\, e^{-\beta F_{dyn}(y)}$$
$$\approx \frac{2\pi}{\sqrt{K_0 K_B}} e^{\beta(F_B + F_{el})} \qquad (10)$$

where $K_B$ is the absolute value of the barrier curvature, and the final approximate relation is accurate when the barrier is beyond of few thermal energy units.

### C. Projected NMCT Force Vertex Behaviors

The dominant length scale(s) of the Fourier-space resolved force correlations (or vertex) in NMCT and the dynamic free energy (per the k-space integration in eqs(6) and (8)) has been established via the so-called "ultra-local" analytic analysis and numerical studies [46,47]. The dynamic mean square force vertex that enters Eq(8) is defined as:

$$V(k) = k^4 C(k)^2 \rho S(k)$$
$$\propto \phi K^4 C(K)^2 S(K) \qquad (11)$$
$$\propto \phi^{-1} K^4 S(K)(1 - S(k)^{-1})^2$$

where in the second and third expressions the wavevectors are non-dimensionalized as $K = k\sigma$. This quantity exhibits (for both hard and sticky spheres) a first and largest peak on the $k \sim k^*$ cage scale; for higher wavevectors the amplitude is a bit lower but with an essentially constant maximum amplitude for all $k > k^*$ (see Fig.SM1a). At the NMCT ideal localization transition the contributions to the dynamic vertex from the $k \sim k^*$ correlations via $S(k \sim k^*)$ dominate for hard and sticky spheres. Since $S(k \sim k^*)$ is a non-monotonic function of attraction strength (see Fig.SM1b), this leads to a non-monotonic cage scale vertex amplitude (Fig.SM1c), resulting in the predicted re-entrant "nose" form of the ideal localization boundary as previously discussed [44,45,47] (also shown in Fig.5 below).

However, well beyond the NMCT crossover volume fraction where the barrier is significant, contributions to the dynamic vertex on scales $k \gg k^*$ become important (since localization is tighter) and eventually dominate, per prior analysis [46,47]. The $k > k^*$ amplitude then sets the energy scale of the caging part of the dynamic free energy, which has been analytically shown [46] to be characterized by a "coupling constant" $\nu \propto \phi g(d)^2$, where $g(d)$ is the contact value of the pair correlation function. For short range attractive forces this coupling constant grows monotonically with attraction strength and decreasing range (see Fig.SM1d) reflecting enhanced local clustering. This trend suggests the non-monotonic re-entrant behavior predicted at the NMCT level may "go away" sufficiently beyond the ideal localization boundary. Numerical studies verify this to be the case [45,47]. Moreover, the non-monotonic variation of $S(k^*)$ with attraction strength, and the vertex amplitude associated it, *weakens* with increasing volume fraction (Figs SM1b and SM1c), a trend consistent with the van der Waals idea that attractions modify structure less as density grows [34,35].

Hence, *if* one accepts that ECNLE theory with the projected vertex can reliably capture basic aspects of activated relaxation, *and* activated motion is essential to explicitly capture in glass, gel and attractive glass forming liquids, then the above behaviors suggest: (i) ad hoc shifting of the ideal MCT boundary into a regime where hopping is important is not justified, and (ii) ECNLE theory based on the force projection approximation will not properly capture non-monotonic dynamic behavior in the high volume regime of Figure 1 relevant to experiments and simulations. Exploring this scenario is the prime goal of our present work.

### III. Projectionless Dynamical Approach

We first analyze Eq(1) based on Newtonian forces. The basic ideas invoked below have a long history in chemical and polymer physics, as previously discussed [42,50,51]. Here we consider a pair potential composed of a singular hard core repulsion plus short range attraction, and how avoiding the projection approximation can be carried out in this context.

## A. Basic Idea

One starts by defining a density field variable, $\tilde{\rho}(\vec{r}, t)$, that differs from the standard 1-body form employed in MCT-like analyzes,

$$\tilde{\rho}(\vec{r}, t) = \sum_j \delta(\vec{r} - \vec{r}_0(t) + \vec{r}_j(t)) \quad (12)$$

This field is the number density of particles a distance $\vec{r}$ *from* the tagged particle at time $t$. Using it, one can formally write Eq(1) exactly as:

$$K(t) = \frac{\beta}{3} \langle \vec{F_0}(0) \cdot \vec{F_0}(t) \rangle$$
$$= \frac{\beta}{3} \int d\vec{r} \int d\vec{r}' \vec{f}(r) \cdot \vec{f}(r') \langle \tilde{\rho}(\vec{r}, 0) \tilde{\rho}(\vec{r}', t) \rangle \quad (13)$$
$$= \frac{\beta}{3} \int d\vec{r} \int d\vec{r}' \vec{f}(r) \cdot \vec{f}(r') g(r) g(r') \Omega(\vec{r}, \vec{r}'; t)$$

where $\vec{f}(r) = -\vec{\nabla} u(r)$ is the bare interparticle force, $\langle \tilde{\rho}(\vec{r}, t) \rangle = \rho g(r)$ and $\Omega(\vec{r}, \vec{r}'; t)$ is a multipoint dynamic correlation function of a "conditional form" defined as,

$$\Omega(\vec{r}, \vec{r}'; t) = \frac{\langle \Delta \tilde{\rho}(\vec{r}, 0) \Delta \tilde{\rho}(\vec{r}', t) \rangle}{\langle \tilde{\rho}(\vec{r}, 0) \rangle \langle \tilde{\rho}(\vec{r}', t) \rangle} \quad (14)$$

$$\Delta \tilde{\rho}(\vec{r}, t) = \tilde{\rho}(\vec{r}, t) - \rho g(r) \quad (15)$$

The complicated function in Eq(14) involves dynamic density fluctuations of a pair of matrix particles *in the vicinity* of the moving tagged particle. Its exact determination is impossible. Schweizer and Chandler [50] and Schweizer [51] previously suggested the simple approximation with a projectionless framework:

$$\Omega(\vec{r}, \vec{r}'; t) \approx$$
$$\rho^{-1} \int d\vec{x} \Gamma_s(\vec{x}, t) S(\vec{r} - \vec{r}' + \vec{x}, t) \quad (16)$$

where, $\Gamma_s$ and $S$ are self and collective van-Hove correlation functions, respectively. This approximation is essentially identical to what is employed in NMCT. Rewriting Eq(13) in Fourier space, the force vertex quantity based on this "projectionless dynamic theory" (PDT) is,

$$\vec{M}_{PDT}(k)$$
$$= \int d\vec{r} \vec{f}(r) g(r) e^{-i\vec{k} \cdot \vec{r}}$$
$$= 4\pi \int_0^\infty r^2 f(r) g(r) \frac{\sin(kr)}{kr} dr \quad (17)$$

This differs from the projection based form of Eq(9), with the bare force now entering weighted by $g(r)$ which quantifies the relative number of neighbors at a separation $r$. Finally, one has:

$$K(t)$$
$$= \frac{\beta \rho}{3} \int \frac{d\vec{q}}{(2\pi)^3} |\vec{M}_{PDT}(q)|^2 S(q) \Gamma_s(q, t) \Gamma_c(q, t) \quad (18)$$

The self and collective Debye-Waller factors are treated the same as in section II.

The general form of the arrested component of the force-force time correlation function in NMCT, and also the dynamic free energy in NLE theory, remain the same as before, *but* with a different force vertex. In the projection approach, the vertex of the dynamic free energy (per Eq(8)) scales as $\sim k^2$ as $k \to 0$, and as $\sim k^{-2}$ as $k \to \infty$. Hence, small wavevector correlations are unimportant. But in PDT the corresponding scaling as $k \to 0$ is $\sim k^0$ since $M_{PDT}$ does not vanish in this limit; at high wavevectors the same $\sim k^{-2}$ scaling is obtained as for the projection approximation. We expect this $k \to 0$ PDT scaling behavior is not correct since the dynamic force correlations (per Figure 2) are highly local, especially for short range attractions. We refer to this as issue (a). A second issue (b) is whether the PDT should be employed for hard spheres. We expect not, since a major argument in favor of the projection is that it renormalizes the singular potential (force undefined) with the nonsingular direct correlation function. These two considerations motivate the "hybrid PDT" approach, as previously sketched [42].

## B. Hybrid Approach

We consider fluids that interact via a pair potential composed of a hard core repulsion plus short range attraction, $V(r)$. In specific applications we employ an exponential form:

$$V(r) = -\epsilon e^{-\frac{r - \sigma}{a}} \quad (19)$$

$$f(r) = -\frac{\epsilon}{a} e^{-\frac{r-\sigma}{a}} \qquad (20)$$

Obviously, as the range changes from $\frac{a}{\sigma} = 0.1$ to $\frac{a}{\sigma} = 0.02$, typical of the soft matter systems of interest, the attractive force increases by a factor of 5, which we expect should have a major consequence on physical bond lifetime, and hence structural dynamics. Given the discussion in section IIIA, we adopt a hybrid approach where the NMCT vertex is retained for the repulsive force (to address issue (b)) and the attractive force is treated with the PDT approximation. One could perhaps view this as in the spirit of equilibrium integral equation theories which construct mixed closures when there are competing repulsive and attractive interactions [36]. But there is a major conceptual difference here since for dynamics one need forces and not potentials, and the range of the attraction will play a stronger and more explicit role.

The hybrid approximation thus separates the total force vertex into repulsive $\vec{M}_R(k) = \vec{M}_{NMCT}(k)$ and attractive $\vec{M}_A(k) = \vec{M}_{PDT}(k)$ contributions as

$$\begin{aligned}\vec{M}_{Hyb}(k) &\approx \vec{M}_R(k) + \vec{M}_A(k) \\ &= \vec{M}_{NMCT}(k) + \vec{M}_{PDT}(k) \\ &= kC_0(k) + 4\pi \int_0^\infty r^2 f(r) g(r) j_0(kr) dr\end{aligned} \qquad (21)$$

where the repulsive force contribution is for the pure hard sphere system. The full structure factor, $S(k)$, still enters per Eq(18). A qualitatively new feature is there is *a negative* cross term associated with correlations or interference between repulsive and attractive forces experienced by a tagged particle. Hence, for *weak* attractions one might expect this cross term will *reduce* dynamical constraints, but for strong enough attractions the net effect will be an enhancement of force correlations relative to the pure hard sphere fluid. This behavior applies at all theoretical levels: NMCT, NLE and ECNLE, which we expect will lead to non-monotonic variation of *all* dynamical properties in a manner *not* tied to being close to the ideal localization boundary. In addition, the consequences of physical clustering in enhancing $g(r)$ near contact and non-monotonic variation of $S(k^*)$ with attraction strength are still present, the key effects embedded in the MCT-like projection approximation for the force vertex discussed in section IIC and SM.

### C. Expected Consequences of the Hybrid PDT Approach

To gain qualitative insight, we crudely estimate how the hybrid force vertex amplitude of Eq(21) behaves. We have verified that the approximate analytic analysis presented below is consistent with our numerical results. We estimate the hybrid-PDT vertex as:

$$|M(k)|^2 \propto \left(\frac{k_B T}{\sigma} g_{HS}(\sigma) - \#\frac{\beta\epsilon}{a} g(a+\sigma)\right)^2 S(k^*) \qquad (22)$$

The first term inside the parentheses qualitatively captures the projection-based behavior for hard spheres (the force scale set by $k_B T/\sigma$) as previously derived [10, 21-24], and the second term describes attractive forces (force scale set by $\epsilon/a$) which grow monotonically with increasing attraction strength *and* decreasing attraction range. Even though $S(k^*)$ is non-monotonic, in the hybrid-PDT approach any "re-entrant" glass melting well beyond the ideal localization boundary will be a combined effect of such a structural correlation effect and interference between attractive and repulsive forces. This scenario suggests the strength of the re-entrancy feature of the ideal localization boundary will be stronger if the hybrid-PDT vertex is adopted compared to its projected analog. Numerical calculations below verify this, and the cross-term is critical for predicting both re-entrant melting and non-monotonic dynamics in the strongly activated regime.

Finally, to address issue (a) discussed at the end of the previous section, we believe the PDT approximation should only be used at the high wavevectors that define the local cage and short range attractions and bond formation. How precisely to implement this idea seems subtle since the *relative* importance of the repulsive versus attractive force contributions to the vertex is crucial in the hybrid PDT approach. Here we adopt a cutoff wavevector, $k_c$, for using the PDT to predict the dynamical consequences of attractive forces. Though we believe this is physically motivated, it is an ad hoc (though minimalist) device. We adopt what we view as the natural choice of $k_c = k^*$, where $k^*$ is the wavevector at the cage peak of $S(k)$. This is an estimate of cage size, which plays a central role in defining dynamical constraints in the NLE and ECNLE theories [10, 19, 21-22]. This cutoff choice is *not* changed based on the system or thermodynamic state. It follows from $S(k)$ and thus varies with volume fraction and attraction strength and range, i.e., $k_c = k^*(\phi, \beta\epsilon, a/\sigma)$. We have verified the numerical consequences of this cutoff for NMCT and ECNLE theory are (inevitably) quantitatively sensitive to its precise value adopted, but predicted trends with control parameters are *not qualitatively* sensitive.

### IV. Theoretical Results

#### A. Ideal Kinetic Arrest Diagram and Localization Length

Figure 4 shows the results of the hybrid PDT based predictions (solid curves) for the ideal localization boundary for 3 short attraction ranges. The classic re-entrant behavior is predicted, with the non-monotonic nose feature becoming stronger as the range decreases. When the attraction range reaches a value typical of thermal liquids ($a \sim 0.5\sigma$), we find (not shown) the re-entrancy feature is essentially gone. The curves for different ranges cross at an intermediate attraction strength, signaling the emergence of dense gel states at lower (but still high) volume fractions. All the hybrid PDT trends are understandable as a consequence of the attractive force scaling as $f \propto \frac{\epsilon}{a}$. This includes curve crossing since 'glass melting' will be stronger via the cross term between repulsive and attractive forces at fixed $\epsilon$ for smaller range, and also the stronger tendency to form an attractive glass at high attraction strengths for a shorter range potential.

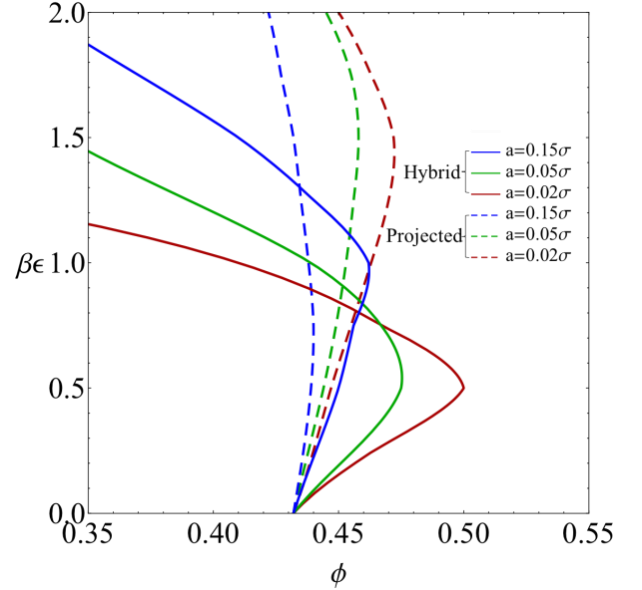

Figure 4: Ideal kinetic arrest map in attraction strength versus volume fraction space based on NMCT with the projected vertex (dashed curves) and the hybrid PDT vertex (solid curves) for three short ranges of exponential attraction.

The corresponding results based on the projected vertex are shown as dashed curves (per prior NMCT studies [44,45,47]). Re-entrant behavior is predicted, though the location and "strength" of the nose feature (degree of non-monotonicity) is weaker than the hybrid PDT results, consistent with the discussion in section IIIC. The ideal NMCT nonergodicity curves for different attractive force ranges also cross but at attraction strengths off scale in Figure 4. For a 'long' range attraction of $a = 0.5\sigma$ the ideal arrest boundary is perfectly vertical (not shown).

Overall, we find *no qualitative* differences between the projected and hybrid PDT force vertex based calculations of the form and trends of the ideal arrest boundaries, but rather that the latter approach just amplifies the trends of the former. Importantly, if the interference cross term in the hybrid PDT is dropped, then *no* re-entrant behavior is predicted. This is consistent with the analytic discussion in section IIIC. Hence, the leading order physical origin for the re-entrant behavior is *different* in the projected and hybrid PDT based implementations of NMCT.

Figure 5 shows analogous projection and hybrid PDT based calculations of the dynamic localization length normalized by its hard sphere fluid value. Both show non-monotonic behavior which is more pronounced with decreasing attraction range. Quantitatively, the hybrid PDT results display a modestly larger effect. Within NMCT the localization length is directly related to the dynamic elastic shear modulus as [44,46] $\beta G' \sigma^3 \propto \phi \left(\frac{\sigma}{r_L}\right)^2$. Hence 're-entrant' behavior for $G'$ is also predicted, which is in qualitative accord with experiments of Pham et al. [52].

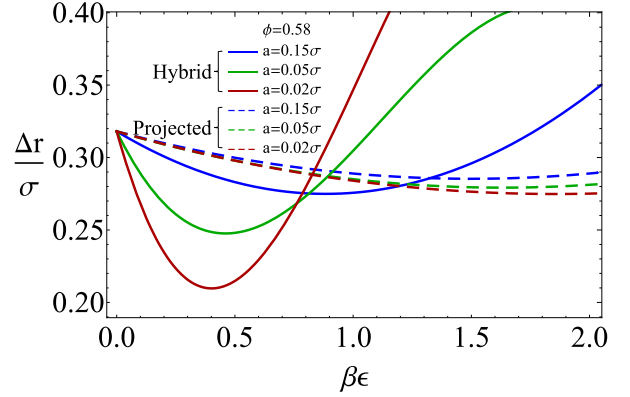

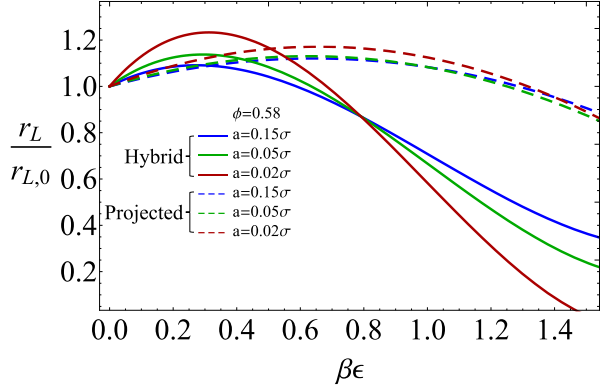

Figure 5: Dimensionless localization length (in units of its hard sphere value) predicted by NMCT theory based on the projected and hybrid PDT vertex approximations at a fixed value of volume fraction as a function of attraction strength for 3 attraction ranges.

### B. Dynamic Free Energy: PDT vs Projected Vertex Results

Going beyond NMCT to implement the activated NLE or ECNLE theories requires several key properties of the dynamic free energy. All are expected to be non-monotonic functions of attraction strength in the hybrid PDT approach. Key quantities include the harmonic curvature $K_0 = \frac{3k_BT}{r_L^2}$ (per Fig.5), the jump distance, and the local and collective elastic barriers.

Figure 6: Non-dimensionalized jump distance $\Delta r = r_B - r_L$ of the dynamic free energy computed based on the hybrid-PDT and projected force vertices at a fixed volume fraction of 0.58 as a function of attraction strength for 3 attraction ranges.

Figure 6 shows representative results for the jump distance as a function of attraction strength for several ranges at a high volume fraction of 0.58. The hybrid PDT theory predicts a strongly non-monotonic variation with a well-defined minimum that is enhanced for shorter range forces, while the projected vertex based analog does not predict these trends. This again emphasizes the central role played by the cross term between repulsive and attractive forces, and the relatively small effect of the non-monotonic change in the cage coherence function, $S(k^*)$, *for* properties related to activated motion. The strong upturn and curve crossing for the hybrid PDT approach mirror those found in Figures 4 and 5, and exist for the same reason.

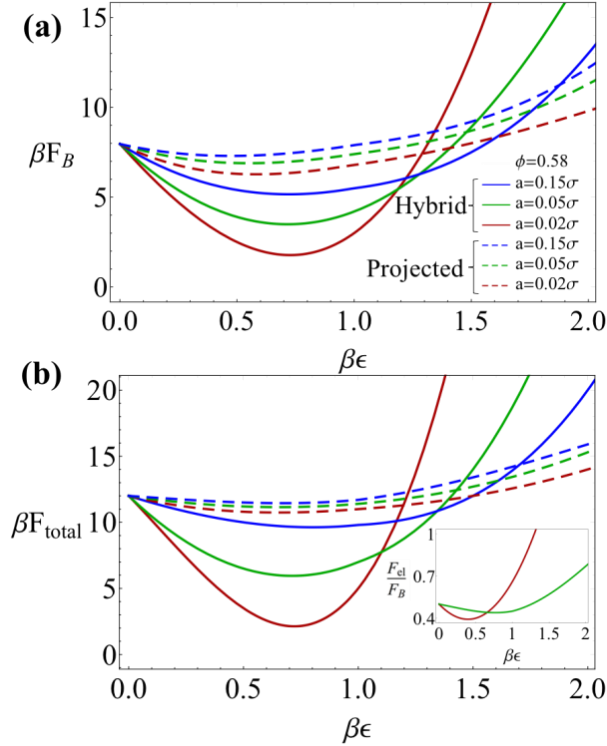

Figure 7: Non-dimensionalized (a) local cage and (b) total barriers for different attraction ranges using hybrid-PDT and projected vertices at a volume fraction of 0.58 as a function of attraction strength for the same 3 ranges in both plots. The inset of panel (b) shows a linear plot of the variation of the *ratio* of the elastic to local cage barriers as a function of attraction strength for dimensionless ranges of 0.02 (red curve) and 0.05 (green curve) based on the hybrid-PDT vertex.

Figure 7 shows analogous results for the local cage and total barriers. The "glass melting" and strongly non-monotonic features, and the trends with attraction range, all mirror those seen in Figures 4-6. The similar behavior of the local and total barriers implies the collective elastic barrier changes in the same qualitative manner with attraction strength and range as its local cage analog. Quantitatively, if $\frac{a}{\sigma} = 0.02$ then the total barrier decreases by $\sim 10 k_B T$ relative to its hard sphere analog, which would produce $\sim 4$ decades faster relaxation. In contrast, but as expected from the discussion in section IIIC, the barriers predicted by the projected vertex, though following the same ordering with range for weak attractions, neither cross at high attraction strength nor show significant re-entrant features. They also grow far more weakly at high attraction strength than the hybrid PDT results. These differences suggest that, although at the ideal localization (dynamic crossover) level the projected and hybrid PDT vertex based theories make qualitatively identical predictions, this is not true in the activated regime.

The inset of Figure 7b shows the ratio of the collective elastic to cage barriers for two attraction ranges based on the hybrid-PDT vertex. This ratio is a key measure of the degree of cooperativity of activated relaxation in ECNLE theory[23,24], and in the bulk correlates with dynamic fragility. Since both barriers are non-monotonic functions of attraction strength, how their ratio behaves is nontrivial. We find a distinct non-monotonic variation of this measure of cooperativity, although the extent of decrease at the minimum is very modest relative to the behavior of the individual barriers. As expected, the overall degree of non-monotonicity is enhanced for shorter range attractive forces.

Taken as a whole, all predicted trends as a function of volume fraction and attraction strength and range qualitatively agree at both the ideal localization and strongly activated levels *within* the hybrid PDT approach. This implies a strong connection between the dynamic crossover (NMCT) and strongly activated (ECNLE theory) physics. One might interpret this as providing modest support for empirical successes claimed by others (for some properties and questions) based on ad hoc shifting of the MCT boundary or using an "effective" larger volume fraction in an ideal MCT calculation. But the validity of such a deduction seems doubtful given trajectories are activated and intermittent at the high volume fractions where glass melting and re-entrancy is observed, physics absent in a shifted ideal MCT scenario.

### C. Activated Structural Relaxation Time

We now assemble all the above results and perform ECNLE theory calculations with the hybrid PDT vertex of the dimensionless mean relaxation time for different attraction ranges and two volume fractions (Figure 8). The calculations assume structural equilibration, which in a practical simulations or experiments may not apply for the largest reduced times shown.

We expect and find similar non-monotonic behavior (including glass melting) for both volume fractions, which are *far* from ideal arrest boundary of $\phi_c = 0.432$ for hard spheres or where the "nose" feature exists in Fig.5. Well beyond the minimum relaxation time, starting around $\beta\epsilon \sim 1.2$, there is a smooth crossover to an attractive glass state characterized by a strong and monotonic growth of the relaxation time due to reinforcing caging and bond formation processes. As expected, both the large speed up of relaxation, and the slower dynamics for strong attractions, are the largest when the attraction range is shortest. On the other hand, for a "long range" attraction typical of thermal (e.g., LJ) liquids ($\frac{a}{\sigma} = 0.5$), one expects the attractive forces are strongly sub-dominant to the repulsive forces at high densities. This is what we find, with the hybrid PDT theory predicting only a slight hint of non-monotonic behavior in Figure 8.

The absolute magnitude of the non-monotonic changes of the alpha relaxation time are dependent on volume fraction, but the behavior is a little subtle. The reason is that as volume fraction increases, the consequences of the excluded volume caging forces grow very strongly, more strongly than their attractive force analog in Eq(22). On the other hand, at higher volume fractions the barriers are larger (the 0.58 hard sphere system has an alpha time ~1000 longer than for 0.55), and hence changes due to attractive forces can induce bigger effects. Figure 8 shows the net effect is the degree of glass melting relaxation time acceleration *grows* with volume fraction, though all qualitative trends with attraction range and the shape of the curves are qualitatively the same. For $\phi = 0.55$ and $\phi = 0.58$, we find $2 - 2.5$ and $4.5 - 5$ orders of magnitude faster dynamics at the most glass-melted state, respectively, for $\frac{a}{\sigma} = 0.02$.

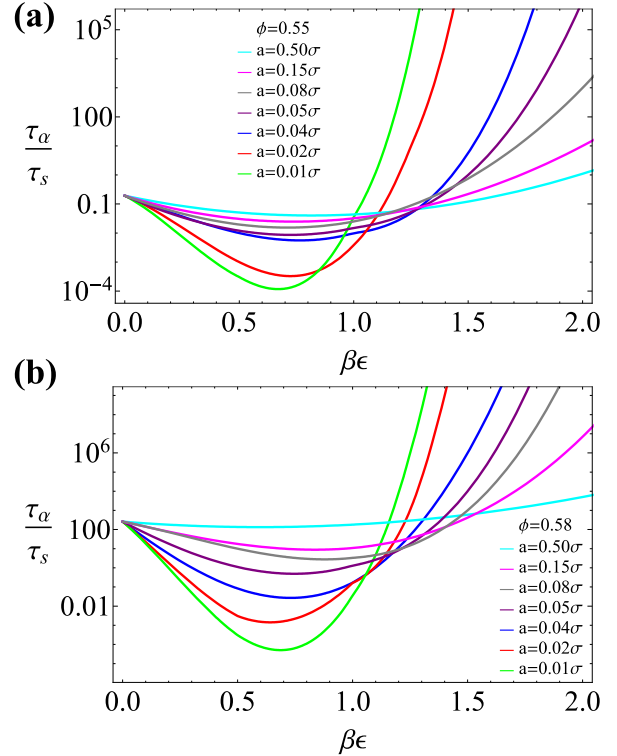

Figure 8: Non-dimensionalized alpha relaxation time as a function of attraction strength for 3 attraction ranges using hybrid-PDT vertex for (a) $\phi = 0.55$ and (b) $\phi = 0.58$.

As the range grows to $\frac{a}{\sigma} = 0.08$, smaller maximum enhancements of relaxation time are predicted, $\approx 1$ and $\approx 1.5 - 2$ decades for the lower and higher volume fractions, respectively. Another trend is that upon increasing attraction strength beyond where the speed up is maximal there is a very strong monotonic growth of $\tau_\alpha$. For example, at $\beta\epsilon = 1.25$, changing $\phi = 0.55 \rightarrow 0.58$ results in a $\approx 3$ decade growth in $\tau_\alpha$ for $\frac{a}{\sigma} = 0.02$, and $\approx 2$ decade growth for $\frac{a}{\sigma} = 0.08$. Even further from the relaxation time minimum (e.g., $\beta\epsilon > 1.5$), larger changes are predicted as strong bonds form and reinforce caging in the attractive glass region. Both the glass melting effect and strong upturn of the relaxation time are consequences of the relative importance of the cross term between strong short range attractive and repulsive forces, as suggested in section IIIC.

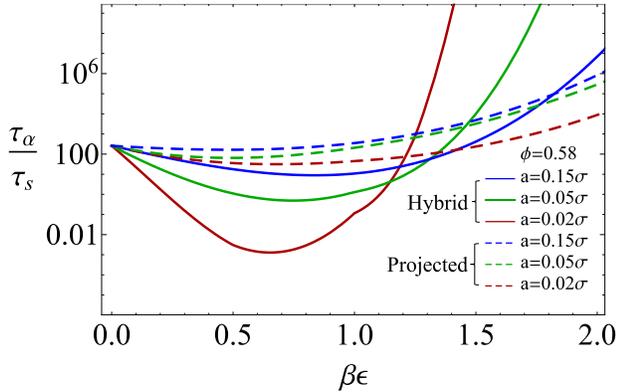

Figure 9: Non-dimensionalized alpha relaxation time at fixed $\phi = 0.58$ as a function of attraction strength for 3 attraction ranges using hybrid-PDT and projected vertices.

Figure 9 compares the hybrid PDT and projection based force vertex ECNLE theory predictions for the dimensionless alpha time at $\phi = 0.58$. As expected from Figure 7, the latter predicts essentially no non-monotonic or glass melting behavior, nor curve crossing at high attractions strengths in the AG regime.

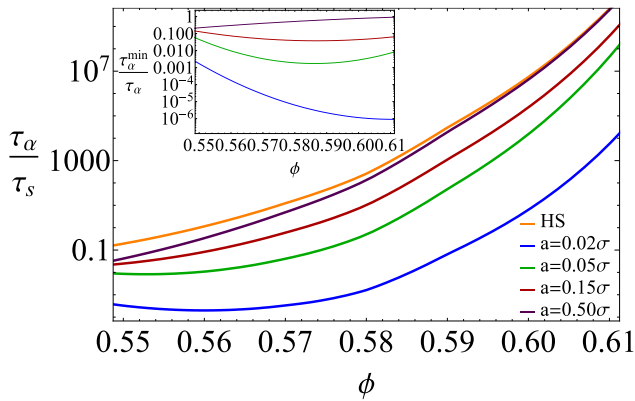

Figure 10: Non-dimensionalized alpha relaxation time at the unique value of attraction strength corresponding to "fastest relaxation" state (maximum glass melting) as a function of volume fraction for different ranges of attraction based on the hybrid PDT vertex. (Inset) maximum change in alpha time as a function of volume fraction for the same ranges of attraction.

Figure 10 collects the hybrid PDT vertex ECNLE theory results from prior figures for the minimum alpha time state (most glass melted, "fastest relaxation"), and supplements them with additional calculations for other volume fractions, to create a summary plot of how the minimum relaxation time changes as a function of attraction range and volume fraction; the pure hard sphere fluid result is also shown (top curve). The log-linear format demonstrates the strong dependence of the magnitude of the maximal glass melting effect on volume fraction. The inset displays the corresponding magnitude of the maximal relaxation time reduction factor.

Broadly, the evolution with volume fraction of the degree of relaxation time speed up displays several regimes. There is only a slight increase if the exponential attraction range is long (0.5), a weakly non-monotonic behavior for intermediate ranges (0.15, 0.05), and strongly decreasing behavior for very short range (e.g., 0.02) though with a tendency for saturation. More detailed trends include: (i) for the shortest attraction range the maximum speed up factor occurs for the higher volume fractions, (ii) as the attraction range becomes longer the maximum change is almost constant for $\phi = 0.575 - 0.61$, (iii) for the most liquid-like system studied (lowest volume fraction) the changes relative to the hard sphere fluid are the smallest.

Finally, we briefly comment on the case of "long range" attractions, which has been studied in recent simulation [38-41] and theoretical [42] work. There the question was differences in activated relaxation time in the lightly supercooled regime between fluids that interact via the continuous WCA repulsion and its analogous LJ potential. Major differences were found in the isochoric simulations which were argued to not be understandable based on any differences in pair structure. The theory of ref.[42] based on a different version of the hybrid PDT ECNLE theory suggested such differences can emerge without any changes of structure in $g(r)$ of the WCA and LJ liquids. Recent machine learning simulations [53] have argued the dynamical differences emerge from small differences in $g(r)$ of the WCA and LJ liquids that are amplified in a specific manner. All of this is beyond the scope of our present work. However, given the prior work on this problem using a different version of the hybrid PDT ECNLE theory [42], this topic is briefly discussed and new calculations presented in the SM. We find differences between relaxation times of WCA and LJ liquids to be influenced by *both* differences in their $g(r)$'s and the direct theoretical treatment of attractive forces.

## V. Comparison with Experiments and Simulations on Attractive Colloids

Given the simplicity of our monodisperse hard sphere plus exponential attraction model, quantitative comparison to colloidal experiments and simulations is not appropriate considering the various complexities of the latter (e.g., polydisperse particle size distribution, solvents, attractive potential form). Rather, we summarize germane studies and their qualitative trends and order of magnitude of the effects. With a couple of exceptions, the studies involved very high volume fractions of $\phi = 0.58 - 0.61$ that are well beyond the ideal nonergodicity transition of MCT or NMCT. In experiments that employ polymer depletion, the range of the effective attraction in units of the colloid diameter was ~ 0.03-0.09. It is difficult (and ill-defined if the functional form of the potential differs) to know the precise values of attraction strength of our exponential model that one should compare with these studies. However, all the experimental and simulation studies observed a strong non-monotonic evolution of the structural relaxation time or diffusion constant with increasing attraction strength, and the existence of a state of maximal degree of speeding up. Within the above range of volume fractions and attraction ranges, our calculations suggest 3 key trends: (i) the maximal degree of speeding up could be as small as 1-2 decades and as large as 1.5-6 decades ($\phi = 0.61$), (ii) stronger rate of increase of the relaxation time past the maximally fluid state than its rate of decrease below it, and (iii) larger non-monotonic variations at higher volume fractions.

Using low molecular weight polystyrene to induce depletion attraction (dimensionless range ~ 0.09) plus high volume fraction (~0.60) polydisperse PMMA hard sphere colloids, Pham et al. [54] discovered a non-monotonic variation of an apparent nonergodicity plateau in the collective dynamic structure factor with increasing strength of attraction. These experiments can measure to time scales up to ~10,000 s, which is only 4-6 decades longer than the elementary Brownian time. If the relaxation process is slower than 10,000 s the system is viewed as "nonergodic". The pure hard sphere suspension showed a nonergodicity plateau of $f(q,\infty) \approx 0.7$, indicating a colloidal glass due to caging constraints. Adding a small amount of polymer only slightly weakened the nonergodic state with $f(q,\infty) \approx 0.62$. Further increasing the attraction strength resulted in a full decay to zero of the time correlation function, the signature of "glass melting" into a fluid. Upon adding more polymer the fluid again became nonergodic, and even more so than for the repulsive glass since $f(q,\infty) \approx 0.95$. These experiments support a re-entrant repulsive glass →fluid →attractive glass scenario. Crude estimates of relaxation time suggest an initial ~2-3 decade speed up of dynamics relative to hard spheres, followed by a very strong increase of relaxation time to well beyond that of the hard sphere system.

Eckert and Bartsch [28] studied a binary mixture of polydisperse crosslinked microspheres of $\phi_{colloid} \approx 0.67$ mixed with non-adsorbing polymer corresponding to a depletion attraction range of $\approx 0.05\sigma$. They observed an arrested→liquid→arrested transition scenario. For no free polymer, $f(q,\infty) \approx 0.89$ per a repulsive glass. For intermediate attraction strengths, $S(q,t)$ decayed to zero per an ergodic liquid. At a higher attraction strength, the system re-arrests with $f(q,\infty) \approx 0.90$. For the ergodic fluids, a structural relaxation time was estimated to be $\approx 3-4$ decade faster than the polymer-free suspension for intermediate attraction strength. For even stronger attraction, a very sharp increase of the relaxation time was found before non-ergodicity again emerges. Overall, the relaxation time varied by 5-6 decades over the observable range.

Recent experiments on the PMMA-PS system were performed [29] and complemented by simulations using an attractive square well potential with a range 3% of the mean particle diameter. The simulations found for $\phi \approx 0.54$ that $\tau_\alpha$ decreases by ~3 decades from its hard sphere value, goes through a minimum, and then even more strongly increases by $\approx 6-7$ decades relative to the minimum. Experiments observed similar behavior. For a higher $\phi \approx 0.59$, simulations found qualitatively the same behavior, but a stronger non-monotonic change of the relaxation time; the difference between the fastest and slowest systems spanned 7-8 decades.

Other experiments on PMMA-PS colloidal suspensions with $\phi \approx 0.58 - 0.60$ and attraction ranges of $\sim 0.05 - 0.08\sigma$ probed the colloid self-diffusion constant [55]. Strong non-monotonic changes were found, with the diffusion constant first rather gently growing with attraction by up to a factor of ~100, followed by a more rapid decrease to a value of order the hard sphere reference. To within the caveat of possible "decoupling" of diffusion and relaxation [2], one expects the diffusion constant scales as the inverse alpha time. Simulations [56] also observed a non-monotonic diffusivity for a short range attraction (dimensionless range ~ 0.03-0.05) at $\phi \approx 0.55$. The behavior is qualitatively the same as the experiments [55], with a ~2 decade increase of diffusion constant, followed by a more rapid ~2 decade decrease at higher attraction strengths.

All of the above studies appear to indicate the same qualitative behavior, including the rough order of magnitude of effects, more rapid growth of the relaxation time at high attraction strength beyond the re-entrant minimum compared to reduction at low attraction strengths, and larger non-monotonic dynamical effects at higher volume fraction. Overall, our calculations are in accord with these trends, and in some cases are quantitatively close.

### VI. Summary and Discussion

We have formulated and implemented a microscopic theory for activated relaxation and glass and attractive glass formation in fluids of hard spheres that interact via strong short range attractions. The approach builds on the ECNLE theory of activated relaxation of hard spheres [23], but reformulates how effective forces are constructed by avoiding the literal "projection" approximation that replaces all microscopic forces by the equilibrium pair correlation function. The hybrid PDT idea explicitly retains information about the bare attractive forces. Interference between repulsive and attractive forces plays a central and novel role, in addition to structural changes due to attraction-induced bond formation and modification of caging. The former effect is the origin of re-entrant dynamical behavior *far* from the ideal localization (dynamic crossover) boundary of NMCT. The theory appears to properly capture the dynamical crossover phenomenology and the strong non-monotonic variation of the structural relaxation time at very high volume fractions as a function of attraction strength and range, consistent with experiments and simulations. ECNLE theory based on the standard projected force vertex does not capture the strong re-entrancy and non-monotonic activated relaxation behavior at very high volume fractions. The overall picture that emerges is qualitatively distinct from phenomenological attempts to use ideal MCT (by shifting the location of the ideal nonergodicity boundary, or ad hoc use of a larger effective volume fraction [57,58]) to understand the slow dynamics of dense suspensions of attractive colloids. The qualitative difference involves not only the central importance of activated hopping, but also how attractive forces create dynamical constraints.

From a theoretical perspective, our approach obviously involves uncontrolled approximations, and its domain of validity is a priori unknown. More specifically, quantification of the hybrid PDT force vertex version of ECNLE theory involves an ad hoc element: a low wave vector cutoff of the dynamical force vertex associated with attractive forces which influences quantitative, but *not* qualitative, aspects of our predictions. The adopted cutoff at $k^*$ seems to us intuitive and consistent with the physical picture of ECNLE theory since it is on the local cage scale. Moreover, the core idea is implemented in a manner that retains predictive power since the cutoff is a priori determined based on the cage peak of $S(k)$ and how it shifts with volume fraction and attractive interaction.

Our approach is immediately applicable to other spherical particle soft matter systems, e.g., colloids with soft repulsions (e.g., microgels, many arm stars) and/or different forms of the short range attraction. It can also be employed to study shorter time physical effects and processes that occur before the activated structural relaxation event. For example, the dynamic plateau shear modulus, or the maximum cage restoring force [11,20,46]. The latter seems closely related to the amplitude of the non-Gaussian parameter [20,59], and is also relevant to yielding [60,61]. Other physical processes on smaller length and time scales, such as the initial stages of activated bond breaking and cage escape, can be addressed. The latter has been recently studied via simulation at an ultra-high volume fraction [30]. Work is in progress in several of these directions.

**Acknowledgements.** This work was supported by DOE-BES under Grant DE-FG02-07ER46471 administered through the Materials Research Laboratory at UIUC.

# Supplementary Material

## A. Projected Dynamic Vertex

Calculations relevant to the discussion in Section IIC are shown in Fig. SM1 below: (a) a representative force vertex per Eq(11) as a function of dimensionless wavevector for a high volume fraction hard sphere fluid (sticky spheres are qualitatively the same), (b) peak of the static structure factor (normalized by hard sphere value) as a function of attraction strength, (c) and (d) as defined in panel (a), the (normalized to the hard sphere) amplitude of the primary cage peak and the $k > k^*$ vertex amplitude, respectively, as a function of attraction strength.

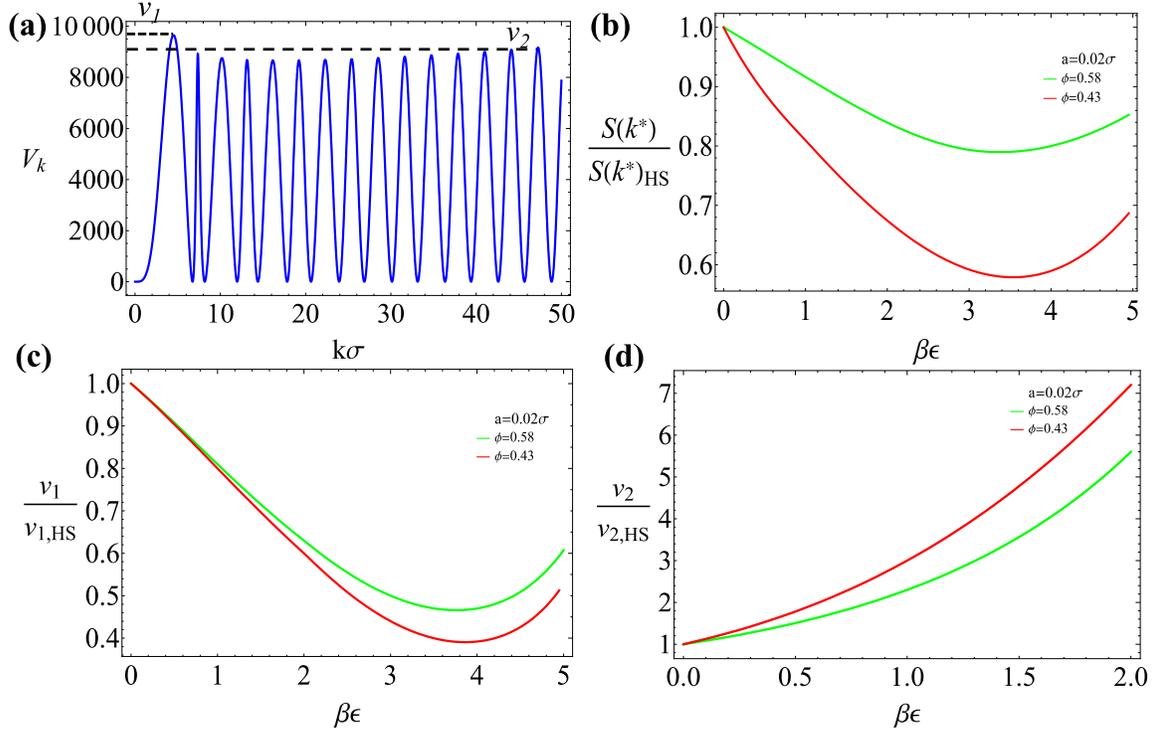

**Fig. SM1.** (a) Dynamic mean square force vertex defined by $V_k = k^4 C_k^2 S_k$ for a hard sphere fluid at $\phi = 0.58$. $v_1$ is the vertex amplitude around the cage peak at $k = k^*$. $v_2$ is the saturated large wavevector vertex amplitude. (b) Dimensionless cage peak amplitude of S(k) as a function of attraction strength for two very different volume fractions at the same range of attraction. (c) and (d) Dimensionless $v_1$ and $v_2$ amplitudes for the same two systems of (b), respectively.

## B. Influence of Long Range Attractions on Activated Relaxation

We briefly elaborate on the question mentioned in section IVC: *what are the differences in activated relaxation of liquids that interact via the repulsive continuous WCA repulsion and its analogous LJ potential?* Three large differences compared to the short range attraction hard sphere fluid model we have studied enter: (i) the WCA repulsion is continuous and involves both density and temperature, (ii) the LJ potential has an effective range that is "long" per a van der Waals liquid, ~ $0.5\sigma$ LJ, and (iii) the attractive force of the LJ potential is zero at its minimum, in contrast to an exponential attraction where the attractive force is a *maximum* at contact.

Here we do not address point (i), but consider a hard core repulsion. We then ask what does our new approach predict if a LJ or "long range" exponential attraction ($a = 0.5\sigma$) is appended to this hard core repulsion (see inset of Fig.SM2). Calculations are performed at fixed volume fraction, in contrast to isochoric simulations [1-4] using the WCA potential where upon cooling the effective hard core diameter, and hence effective volume fraction, grows. We are not advocating any precise connection to the prior simulation studies, but do expect our results are qualitatively relevant for the following four questions. (a) Do long range attractions at high densities matter for the alpha relaxation time? (b) Does using a LJ versus exponential form of attraction matter? (c) Are attraction-induced changes of $g(r)$ important? (d) What is the effect of increasing volume fraction?

For discussion purposes, Figure SM2 shows simulation results of Berthier and Tarjus [1-4] re-plotted in the format of the ratio of the relaxation time of the LJ fluid to its WCA analog as a function of dimensionless attraction strength for 4 values of mixture total number density. One sees attractions have a large influence. Crucially, note that at *fixed* attraction strength, a higher density *enhances* the effect, a trend perhaps not obvious from plots in the original publications since the range of reduced attraction strengths simulated varied enormously with density.

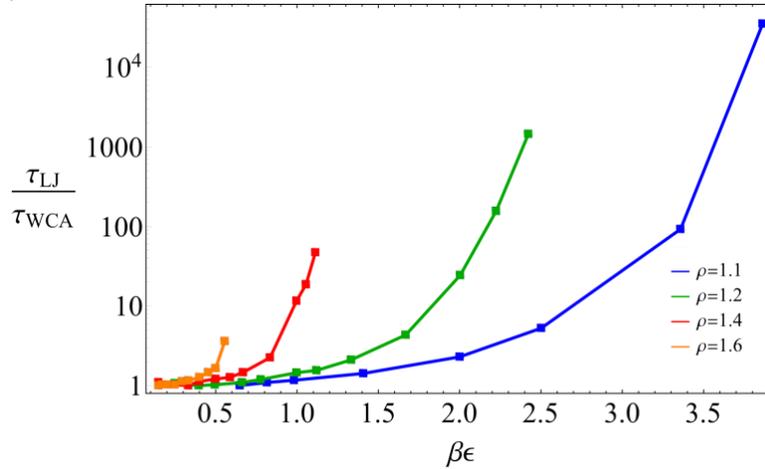

**Fig. SM2.** Ratio of the mean alpha relaxation time of binary sphere mixtures based on either the LJ or WCA pair potentials as obtained in the simulations of ref. [1].

We employ PY theory [5] to compute $g(r)$ and $S(k)$, the quantitative accuracy of which is unclear, but we expect qualitative trends are sensible. Results are shown only for the PDT hybrid vertex version of ECNLE theory, which includes the cross or interference term between repulsive and attractive forces (*ignored* in ref.[6] for the WCA and LJ systems). The "full" $g(r)$ including changes due to the attractive potential are included in Eq(21) unless otherwise stated.

The main frame of Figure SM3 shows the theoretical relaxation time ratio at packing fractions of 0.55 and 0.58. Consider first the LJ attraction model results. For dimensionless attraction strengths up to unity, the hard sphere relaxation time is essentially *unaffected* to within 10-15% or less. At higher attraction strengths, there is a modest growth of the ratio, which is much stronger at the higher packing fraction. The latter trend is in qualitative accord with simulation results of Fig.SM2.

We also checked the influence on the relaxation time of the weak clustering in $g(r)$ induced by the LJ attraction in the PDT vertex of Eq(21). For example, at $\beta\varepsilon = 2$, the contact value of $g(r)$ is ~9 and 7 at $\phi = 0.58$ and $\phi = 0.55$, respectively, compared to the hard sphere fluid values of ~6.8 and ~5.7. We find these attraction-induced changes of local structure *do* have a significant effect on the relaxation time (plot not shown). For example, at $\beta\varepsilon = 2$, the time ratios if these structural changes are ignored are very low, ~ 1.5 and 2 at the two volume fractions studied, compared to ratios of ~ 3.5 and 8 if they are included as seen in Fig.SM3. Our deduction that changes of $g(r)$ when a LJ attraction is introduced might be the leading order effect in determining changes of dynamics (when implemented in the PDT hybrid vertex ECNLE theory framework) is perhaps qualitatively consistent with the recent machine learning study [7]. We note our findings conflict to some degree with those in ref[6] which also were based on the hybrid PDT ECNLE theory since that work (i) ignored the cross term between attractive and repulsive forces for the LJ potential, and (ii) incorrectly employed the $j_1(x)$ Bessel function in Eq(17) instead of the correct $j_0(x)$ function in numerical calculations.

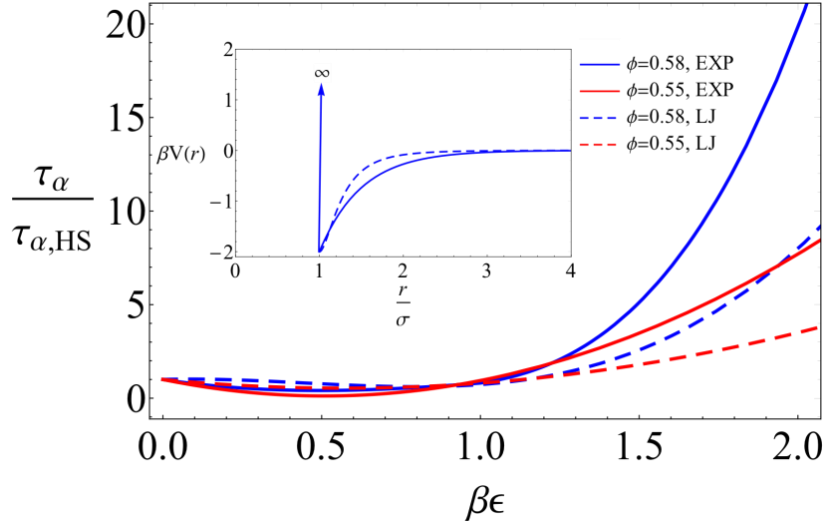

**Fig. SM3.** Ratio of the theoretical alpha relaxation time to its hard sphere analog as a function of attraction strength for two volume fractions. Results are shown for the LJ and exponential models of an attractive tail. Inset shows a schematic of the two potentials (solid is an exponential with $a = \sigma/2$ and dashed is a shifted LJ attraction).

Finally, consider the results in Fig.SM3 based on the long range exponential attraction. The exponential attraction induces larger and more rapidly growing with attraction strength changes of the relaxation time ratio. We find this to be physically intuitive since the attractive force is a *maximum* at contact for an exponential potential, but *zero* for the LJ potential, a key difference which explicitly enters the hybrid PDT vertex of Eq(21). We suggest new simulations be performed to establish the role of the functional form of the attractive force near contact. Concerning attraction-induced changes of $g(r)$ on the relaxation time, the situation is qualitatively identical to what we found for the LJ potential discussed above. For example, if attraction-induced changes of $g(r)$ are ignored in Eq(21), then the relaxation time ratio drops from ~ 22 and 8 at $\beta\varepsilon = 2$ and $\phi = 0.58\ and\ 0.55$, respectively, to ~ 13 and 2.